\documentclass[12pt,preprint]{aastex}

\newcommand{\sgra}{Sgr A$\textrm{*}$}

\renewcommand {\deg}   {\mbox{$^\circ$}}

\newcommand   {\arcs}  {\mbox{$^{\prime\prime}$}}

\newcommand   {\kms}   {\mbox{km\,s$^{-1}$}}
\renewcommand {\ga}    {\mbox{\rlap{\hbox{\lower5pt\hbox{$\sim$}}}\hbox{$>$}}}
\renewcommand {\la}    {\mbox{\rlap{\hbox{\lower5pt\hbox{$\sim$}}}\hbox{$<$}}}

\begin{document}
\def\kms {\hbox{km{\hskip0.1em}s$^{-1}$}} 
\voffset=-0.8in

\def\msol{\hbox{$\hbox{M}_\odot$}}
\def\lsol{\hbox{$\hbox{L}_\odot$}}
\def\kms{km s$^{-1}$}
\def\Blos{B$_{\rm los}$}
\def\etal   {{\it et al.}}                     
\def\psec           {$.\negthinspace^{s}$}
\def\pasec          {$.\negthinspace^{\prime\prime}$}
\def\pdeg           {$.\kern-.25em ^{^\circ}$}
\def\degree{\ifmmode{^\circ} \else{$^\circ$}\fi}
\def\ut #1 #2 { \, \textrm{#1}^{#2}} 
\def\u #1 { \, \textrm{#1}}          
\def\nH {n_\mathrm{H}}
\def\ddeg   {\hbox{$.\!\!^\circ$}}              
\def\deg    {$^{\circ}$}                        
\def\le     {$\leq$}                            
\def\sec    {$^{\rm s}$}                        
\def\msol   {\hbox{$M_\odot$}}                  
\def\i      {\hbox{\it I}}                      
\def\v      {\hbox{\it V}}                      
\def\dasec  {\hbox{$.\!\!^{\prime\prime}$}}     
\def\asec   {$^{\prime\prime}$}                 
\def\dasec  {\hbox{$.\!\!^{\prime\prime}$}}     
\def\dsec   {\hbox{$.\!\!^{\rm s}$}}            
\def\min    {$^{\rm m}$}                        
\def\hour   {$^{\rm h}$}                        
\def\amin   {$^{\prime}$}                       
\def\lsol{\, \hbox{$\hbox{L}_\odot$}}
\def\sec    {$^{\rm s}$}                        
\def\etal   {{\it et al.}}                     
\def\la{\lower.4ex\hbox{$\;\buildrel <\over{\scriptstyle\sim}\;$}}
\def\ga{\lower.4ex\hbox{$\;\buildrel >\over{\scriptstyle\sim}\;$}}
\def\refitem{\par\noindent\hangindent\parindent}
\oddsidemargin = 0pt \topmargin = 0pt \hoffset = 0mm \voffset = -17mm
\textwidth = 160mm  \textheight = 244mm
\parindent 0pt
\parskip 5pt

\shorttitle{Sgr A*}
\shortauthors{}

\title{Radio Continuum Observations of the Galactic Center:\\
 Photoevaporative Proplyd-like Objects near Sgr A*}
\author{F. Yusef-Zadeh$^1$, D. A. Roberts$^1$, M.  Wardle$^2$, 
W. Cotton$^3$, R. Sch\"odel$^4$ \& M. J. Royster$^1$
}
\affil{$^1$Department of Physics and Astronomy and CIERA, Northwestern University, Evanston, IL 60208}
\affil{$^2$Department of Physics and Astronomy, Macquarie University, Sydney NSW 2109, Australia}
\affil{$^3$National Radio Astronomy Observatory,  Charlottesville, VA 22903}
\affil{$^4$Instituto de Astfisica de Andalucia (CSIC), 
Glorieta de la Astronomia S/N, 18008 Granada, Spain}

\begin{abstract} 
We  present radio images  within 30$''$ of Sgr A* based   on recent 
VLA observations at  34 GHz with  7.8 $\mu$Jy  
sensitivity and resolution  $\sim88\times46$ milliarcseconds (mas). 
We report 44 partially resolved compact sources 
clustered in two regions in the E arm of ionized gas that  orbits Sgr A*.  
These  sources have  size scales 
ranging between  $\sim$ 50  and  200 mas (400 to 1600 AUs), 
and a  bow-shock  appearance facing 
the direction of Sgr A*. 
Unlike the  bow-shock sources previously identified in the near-IR
but associated with massive stars,  
 these 34 GHz sources do not appear to have near-IR counterparts at 3.8 $\mu$m. 
We interpret these sources  as  a candidate population of 
photoevaporative protoplanetary disks (proplyds) that are associated with newly formed low mass stars
with mass loss rates  $\sim10^{-7} - 10^{-6}$ \msol\, yr$^{-1}$ 
and are located at the edge of a molecular  cloud outlined by ionized gas. 
The disks  are externally illuminated by  strong Lyman continuum 
radiation from
the $\sim$100 OB and WR massive stars distributed within  10$''$ of Sgr A*. 
The presence of proplyds implies current in-situ star formation activity near Sgr A* and 
opens a window  for the first time  
to study low mass star, planetary and brown dwarf formations  near a supermassive black hole. 
\end{abstract}

\keywords{Galaxy: center - clouds - ISM: general - ISM - radio continuum - stars: formation - protostars}

\section{Introduction}

The Galactic center hosts a young population of stars centered on the strong radio source Sgr A* 
which coincides with a 4$\times10^6$ \msol\, black hole (Reid and Brunthaler 2004; Ghez et al.  2008; 
Gillessen et al. 2009). The young stellar cluster lies  within 1 and 10$''$ (0.039--0.39 pc) 
of Sgr A* and 
consists of about one hundred young massive OB and WR stars 
(Paumard et al.  2006; Lu et al.  2009).

The region within 20$''$ of Sgr A* shows a number of infrared excess sources (Viehmann et 
al. 2006; Muzic et al. 2008; Eckart et al. 2013). Some of these dusty sources are hot, young 
stars interacting with the interstellar medium and creating bow-shock structures. 
Another group of 
infrared excess stars have partially resolved radio 
continuum counterparts and  are interpreted as massive young stellar objects (YSOs) irradiated 
by the strong UV emission from the central stellar cluster (Yusef-Zadeh et al. 2014). Another 
indicator of star formation is the  SiO (5-4) line emission from a dozen unresolved 
sources within 0.5 pc of Sgr A* (Yusef-Zadeh et al. 2013). These may be  
highly embedded protostellar outflows, signifying an early phase of star formation near Sgr A* 
in the last $10^4-10^5$ years (Yusef-Zadeh et al. 2013).

A number of recent studies support in-situ star formation near Sgr A* as  the origin of massive 
stars in the last few million years (e.g., Genzel et al.  2010).  
In the disk of the Galaxy, circumstellar disks associated with low mass stars are expected to survive 
their hostile massive star formation environments. 
A number of circumstellar disks associated with   low mass stars 
show cometary
appearance   and have been identified 
as proplyds (protoplanetary disks) 
in young star forming regions such as the Orion Nebula Cluster (ONC; O'Dell \& Wen 1994),  Carina Nebula (Smith et al. 
2005) and the Trifid Nebula  (Yusef-Zadeh et al. 2005).  
The  proplyds  of the ONC 
are photoevaporated 
and photoionized by the UV radiation from $\theta^1$ Ori C  and generally exhibit 
bow shock structures
(Johnstone, Hollenback \& Bally 1998; 
St\"orzer \& Hollenbach 1998).

We present radio continuum observations within 30$''$ of Sgr A* and report the detection of 44 
radio continuum sources in 2 groupings 15$''$ and 20$''$ NE of Sgr A*. Our 34 GHz observations reveal 
partially resolved sources with typical sizes of a few hundred AU at the 8 kpc distance to the Galactic 
center.  In analogy with the proplyds found in the ONC, 
 we interpret these radio continuum 
sources as proplyd candidates that are photoionized and photoevaporated by the radiation emitted by 
massive OB and WR stars near Sgr A*. The upper limits to the infrared flux from the Galactic center 
proplyd candidates are consistent with gaseous disks orbiting low mass stars at a projected distance of 
$0.6-0.8$ pc from Sgr A*.

\section{Observations and Data Reduction}

Radio continuum observations were   carried out with  the Karl G. Jansky Very Large Array       
(VLA)\footnote{Karl G. Jansky Very Large Array (VLA) of the National Radio
Astronomy Observatory is a facility of the National
Science Foundation, operated under a cooperative agreement by Associated Universities, Inc.}
in its  A-configuration at 34.5  GHz on March 9, 2014 (14A-232). 
We observed  a field of view with a radius of 39$''$ centered on   Sgr A* in 
Ka-band (8.7 mm) 
using the 3-bit sampler system, which provided full polarization correlations in 4 
basebands, each 2 GHz wide, centered at 
31.5, 33.5, 35.5 and 37.5 GHz.
Each baseband was composed of 16 
subbands  128 MHz wide.  Each subband was comprised of 64 channels,  each   2 MHz wide.  
We used 3C286 to 
calibrate the flux density scale and used both 3C286 and J1733-1304 
(aka NRAO530) to calibrate the bandpass.  We used 
J1744-3116 to calibrate the complex gains.  
The constructed image    is of the 30\arcs\ surrounding \sgra\, 
and has  a spatial resolution of 
$\sim88\times46$ milliarcseconds (mas)
(PA$=-1.6^\circ$) 
and  dynamic range   $\sim2.2\times10^5$.
Using the large 8 GHz band, the 1$\sigma$ rms sensitivity is 7.8 $\mu$Jy,  which is a significant improvement 
over older observations which had only 100 MHz of bandwidth.
 


L$'$-band (3.8 $\mu$m) data were taken with the Nasmyth Adaptive Optics System (NACO)
at the Very Large Telescope (VLT; program 089.B-0503(B)) 
between June 26 and September 6, 
2012 with the NACO L27 camera (pixel scale 27 mas pixel$^{-1}$). 
A mosaic of about $50''\times50''$ size was 
constructed  using five  pointings with  total exposure time ranging 
between $\sim$3000 and 5500s. 
The images were reduced with the speckle holography method, as described 
in detail in Sch\"odel  et al. (2010, 2013). 
 


We also present an SiO (5-4) spectrum derived from  
Atacama Large Millimeter Array (ALMA) observations 
carried out by the Science Verification team.
Details  of these observations are given in Yusef-Zadeh et al. (2013). 


\section{Results}

Figure 1a shows a 35$''\times23''$ region of ionized gas associated with the E and N arms of Sgr A West, 
the mini-spiral HII region at 34 GHz. This spectacular image reveals a variety of new compact and 
diffuse sources: radio counterparts to near-IR identified massive stars such as IRS 16NE (Yusef-Zadeh et 
al. 2014), a ridge of blob-like structures arising from Sgr A* extending toward the SW (Yusef-Zadeh et 
al. 2013), a spherical shell-like structure surrounding the M2 supergiant star IRS 7 (Serabyn et al. 
1991), radio emission from the dusty AGB star IRS 3 (Pott et al. 2005) or 
from an embedded WR star (Horribin et al. 2004),   the magnetar SGR J1745-29 (Kennea et 
al. 2013) and a large number of isolated and clustered compact HII regions. 

We focus on the two concentrations of partially resolved 
sources denoted  RC1 and RC2, which appear to be associated with the E arm of the mini-spiral HII 
region.  Figure 1b shows the region in the E arm where
44 radio sources are 
detected.  Figure 1c shows labeled sources P1--P44. The brightest sources,
with peak flux densities ranging between 
31  and 482 $\mu$Jy,  are mainly distributed in RC1  within  a 8$''\times5''$  region of the 
E arm,  centered 
roughly 15.2$''$ (0.61 pc) NE of Sgr A* at  a position angle of 82$\pm8^\circ$.
There are  also   weaker sources in  RC2,  P40--P44, with peak flux densities ranging between 
34 to 53 $\mu$Jy and centered $\sim20''$ from Sgr A*.  
The 34 GHz integrated flux densities of these partially resolved  compact sources range 
between 29  and 1340  $\mu$Jy per 88$\times44$ mas beam. 
We note that the E arm in Figure 1a consists of two elongated ionized structures running parallel to each other 
forming  an  elliptical loop-like structure.
The collection of radio sources are mainly found on the northern edge 
of the E arm. 
Table 1 gives the ID number of 44 detected radio sources (P1--P44)
coordinates, offsets from Sgr A*
in increasing angular distance, 
the total positional accuracy, peak and integrated fluxes at 34 GHz  
in Columns 1 to 7, respectively.  

To illustrate the morphology of the newly resolved radio sources, Figure 2a shows the region of 
the highest concentration of 
proplyd candidates RC1. 
Almost all the sources have  bow shock  or elongated structures. 
Images of two  candidates,  P26 and P8 (Table 1), are presented in Figure 2b,c, 
respectively.  Typical  widths of  bow-shock structures are $\sim$45 mas and the apices of the emission 
from the bow shocks 
are  $\sim$100 mas corresponding to $\sim$800 AU.  
The bows  point to
within a few arcseconds of Sgr A* where young massive stars are concentrated. 
This  morphology  strongly suggests 
 that the resolved sources are externally illuminated by the central cluster of massive stars. The diffuse 
emission from the southern edge of the E arm (Fig. 1a) has a similar appearance suggesting that the 
massive stars are responsible for ionizing both the diffuse and compact sources in Sgr A West. The bow 
shock sources in the diffuse E arm are evaporating gaseous globule (EGG) candidates (see Fig. 1a) after 
Hester et al. (1996) and are an order of magnitude larger than those in RC1 and RC2.


The newly detected radio sources in the Galactic center show similar morphologies to the proplyds found in star 
forming regions in the Galaxy. Radio continuum observations of M42 in 
Orion first detected free-free emission from 
resolved sources with electron densities of $\sim10^6$ cm$^{-3}$ (Churchwell et al. 1987; Garay, Moran \& Reid 1987). 
HST observations of these sources, dubbed proplyds, showed cometary structures with bright heads pointing toward 
$\theta^1$ Ori C (O'Dell \& Wen 1994). We interpret the ionized gas in P1--P44 as arising from photoevaporated 
circumstellar disks that are photoionized by strong sources of UV radiation in the direction of Sgr A*.  The nuclear 
wind from the cluster of massive stars within few arcseconds of Sgr A* could also contribute to the asymmetric shape 
of these proplyd candidates (Serabyn et al. 1991).  The EUV Lyman continuum ionization radiation from massive stars 
is estimated to be $\Phi\approx 2.5\times10^{50}$\,s$^{-1}$ (Genzel, Hollenbach \& Townes 1994) and for an assumed 
$\sim 1$ pc distance from the source of ionization, the incident ionizing photon flux is
$\Phi/(4\pi\,(\mathrm{1\,pc})^2) \approx 2\times10^{12}$\,s$^{-1}$\,cm$^{-2}$.
This  is similar to that 
in Orion where the  proplyds in M42 are located $\sim4\times10^{17}$ cm 
from $\theta^1$ Ori C with $\Phi\approx10^{49}$ s$^{-1}$. 
Furthermore, the brightest sources  in RC1 are closest   to the central cluster  and 
the typical peak flux density  of radio sources in RC1 is stronger than  
that of  RC2. 
This is  consistent with the suggestion that both RC1 and RC2 are 
photoionized by the same stellar cluster  near Sgr A* and that 
RC1 is physically closer to Sgr A* than RC2. 

Assuming a spherical source with solid angle corresponding to a radius of 500 AU and 
electron density filling the sphere with volume filling factor of 1, the total mass of ionized gas $M_i$
and the electron density  
are estimated to be 2.5$\times 10^{-4}$ \msol\, and  $1.2\times10^{5}$ cm$^{-3}$, respectively. 
The electron density  increases if we use a flattened geometry for  the shape of ionized sources 
and a smaller volume filling  factor. 
Assuming that 
the ionized gas expands at the sound speed ($c_s\approx 10$\,km\,s$^{-1}$ for $T=8\,000$\,K), the 
mass-loss rates due to photoevaporation for sources with flux densities between 0.1 and 1 mJy 
are  $M_i/(r/c_s)\approx 10^{-7} - 10^{-6}\, 
\msol\, \rm yr^{-1}$, respectively, 
where $r$ is the radius of the disk. This must be replenished on the expansion time scale $r/c_s \approx 
240$\,yr, implying the existence of a reservoir of neutral material, most likely a disk associated with 
a low mass star. If the source with a mass loss rate 
$10^{-6} \msol\, \rm yr^{-1}$
is 10$^{4}$ years old, then  1$\times10^{-2}$  \msol\, has been  lost today, 
comparable to the disk masses around YSOs. 

In principle, the source of material could be an isolated molecular globule but we can place improbable constraints 
on its mass $M$, as follows. First, the clump must be sufficiently bound by self gravity to avoid disruption by the 
tidal field of Sgr A*. At distance $d$ from Sgr A*, this implies that its density exceeds the Roche density, i.e. $\nH 
\ga 10^8\,(d/\u pc )^{-3} \ut cm -3 $.  Second, models of photoevaporative flows (e.g., St\"orzer \& Hollenbach 1990) 
show that heating by FUV photons drives a neutral wind from the surface that reaches escape speed before being ionized 
by EUV photons at a somewhat larger radii.  Thus the neutral material passes through a sonic point at some radius $r \la 
500$ AU.  At this point the flow speed, sound speed, and escape velocity $\sqrt{2GM/r}$ are all equal, so we can 
obtain an estimate for the total mass inside the sonic surface by noting that the sound speed in the warm atomic wind 
will be $\sim 3 \u km \ut s -1 $ (corresponding to $\sim 1500$\,K), yielding $M \approx 0.5 \, r_{100} \, \msol$, and 
mean density inside the sonic point $<\nH> \, \approx 3\times 10^{10}\,r_{100}^{-2}$ cm$^{-3}$, where $r_{100} = 
r/(100\u AU ) \la 5$.  As  such a dense and massive globule would collapse and form a YSO, it
is  therefore much more plausible that the reservoir of neutral material is instead a circumstellar disk, with $M$ being 
dominated by the central star.

There is further evidence for current  star formation activity in the E arm from ALMA observations which 
detected 11 unresolved sources of SiO (5-4) line emission within 0.5 pc of Sgr A* 
(Yusef-Zadeh et al. 2013). The SiO sources were interpreted as highly embedded 
protostellar outflows with typical mass of swept up molecular gas $\sim0.2$ \msol.  We searched for 
additional SiO emission from the data published earlier 
(Yusef-Zadeh et al. 2013) 
and found a weak SiO source 12 toward the cluster of proplyds. Figure 3a 
shows the positions of SiO (5-4) sources  4 and 5 (Yusef-Zadeh et al. (2013) and the new 
source  12, drawn as ellipses, superimposed on a grayscale continuum image st 34 GHz. Figure 
3b shows the spectrum of the new SiO source 12 
at $(\alpha, \delta)= 17^h 45^m 41^s.36, -29^{\circ} 0' 24''.57$. 
A Gaussian fit to  this spectrum gives peak flux density 13.69$\pm$1.81 mJy beam$^{-1}$ corresponding to 
SiO luminosity $\sim4\times10^{-6}$ \lsol. 
We note that the SiO (5-4) sources  4, 5 and  12  
coincide with the concentration of proplyds candidates. 
The parameters of the fitted spectra of  SiO (5-4) sources  4, 5 and 12 
show central velocities  $\sim$ 8,  14  and 36 \kms\, 
and total linewidths of $\sim25$, 18 and 54 \kms\, respectively. 
These characteristics are very similar to those found in low mass protostellar outflows found in 
the Galactic disk (Gibb et al. 2004; Yusef-Zadeh et al. 2013). 
The presence of proplyd candidates in the vicinity 
of SiO (5-4) source is consistent with this interpretation, signifying 
that low mass star formation near Sgr A* has occurred within the last $10^4-10^5$ years.

We also searched for near-IR  counterparts to the radio sources and placed a 1$\sigma$  upper limit of 
8.6 $\mu$Jy at 3.8 $\mu$m for most sources. 
This population  differs from 
the recently reported 44 GHz sources (Yusef-Zadeh et al. 2014) 
and the near-IR identified bow-shock sources  (Tanner et al. 2005; Sanchez-Bermudez  et al. 2014)
in that they do not have bright near-IR counterparts. 
We can not find  near-IR stellar counterparts to radio sources listed in Table 1
with certainty  because of the lack of stellar proper motion data and confusing sources in this region.  
Two exceptions are 
proplyd candidates P7 and P8 which appear to have  weak 3.8 $\mu$m counterparts at a level of 
$\sim20\,  \mu$Jy pixel$^{-1}$. 
The near-IR luminosity  
is estimated to be 
4$\times10^{-2}$ \lsol\, using  4 $\pi d^2 \times \nu\, S_{\nu}$ 
where the flux density $S_{\nu} \sim 3\sigma$ from a pixel size of $\sim200$ AU (25 mas).  
Most of the energy from an embedded star and the disk 
emerges in the far-IR  and millimeter parts of the spectrum. 
Future high resolution observations of these sources should
provide better constraints on the luminosity of the proplyd candidates. 

\section{Discussion}

We report the detection of 44 partially resolved radio continuum sources with typical sizes of $\sim$300-1000 AU. 
We interpret the newly detected cluster of radio sources as a population of 
proplyd-like objects, implying that low mass star formation is presently taking place in the Galactic center.

First, the radio sources display a bow-shock morphology 
facing the  luminous stars located at the Galactic center. 
The second line of argument is the lack of luminous near-IR counterparts to radio sources is consistent 
with an embedded low mass star.  Like other star forming regions in the Galaxy where proplyds are 
detected with tear-shaped geometry or a disk silhouetted against the background radiation, the emission 
from radio proplyds as well the shape and size of proplyds are consistent with being photoevaporated 
from a gaseous disk by the FUV radiation and then photoionized by the EUV Lyman continuum radiation from 
hot stars in the inner 0.4 pc of the Galaxy (Johnstone et al. 1998; St\"orzer and Hollenbach 1999). 
In addition, there is a lack of correlation between the luminosity and velocity linewidths of SiO (5-4) line 
emission from the region where the cluster of proplyd-like objects (RC1) is detected. The relationship 
is consistent with that found in low mass star formation  in the disk of the Galaxy. 

Third the proplyd candidates lie at the edge of  an elongated molecular cloud
which provides a reservoir of gas that feeds  star formation activity. 
The morphology 
of ionized gas in the E arm (Fig. 1a) suggests that an elongated molecular cloud with an extent of 
20$''\times5''$ lies within the inner pc of the Sgr A*. Proplyd candidates are likely produced from this 
cloud. Support for this suggestion comes from the presence of a radio dark cloud (RDC) associated with 
the E arm. The dark elongated feature in Fig. 1 is completely surrounded by the ionized gas associated 
with the E arm. A neutral gas cloud embedded within a bath of radiation field of the Galactic center 
suppresses radio continuum emission from the interior of the cloud but is edge brightened.  The dark 
feature within the E arm is anti-correlated  with  H$_2$ 
emission detected at 2.12 $\mu$m (Fig. 5f of Yusef-Zadeh 2012). There is also the appearance of 
a small dark feature near RC2 but needs to be confirmed in future molecular line observations. 
The presence of radio dark clouds   implies  that there 
is a supply of dense gas  where the proplyds candidates are detected, thus feeding star formation 
activity near Sgr A*. Figure 4 shows a schematic diagram of the radio cloud sandwiched between the 
ionized gas associated with the E arm and the cluster of proplyd-like objects (RC1) formed at the 
northern edge of the cloud. 
The large-scale bow-shock  sources to the south of the E arm (Fig. 1a)
could belong to the class of 
EGGs similar to 
sources seen in M16 (e.g., Hester et al. 1996).

In order for gas clouds near Sgr A* to collapse and form stars, two conditions must be met. One is that 
self-gravity must overcome the strong tidal shear of Sgr A* such that the density of the gas n$_{\rm H}$
 has to be 
greater than $2\times 10^8 (\rm d/1 pc)^{-3}$ cm$^{-3}$. The other is that the 
cloud mass 
exceeds the Jeans mass $M_J = 1.5\, (T/100 K)^{3/2}\, (n/10^8\, cm^{-3})^{-1/2}$ \msol. 
The combination of these two conditions give $M_J > 1\, 
(d/1 pc)^{3/2}\, (T/100 K)^{1.5}$ \msol. 
Assuming that the gas density is sufficiently high, this suggests that 
the low mass stars 
are preferentially 
formed within a few acrseconds of  Sgr A* with the implication that the 
IMF may be bottom heavy. 
The young stars with $5''-10''$  of Sgr A* 
are reported to have a top-heavy IMF (Bartko et al. 2010, Lu et al. 2013). 
If  clumps with Jeans mass contract and sub-fragment, then 
a population of brown  dwarfs  are expected to lie close to Sgr A*.
An additional  implication of low mass star formation near Sgr A* is that 
proplyd-like objects in the FUV dominated region 
have lower mass-loss rates, perhaps could retain sufficient disk mass 
to allow potential planetary formation (Mann et al. 2014).

In conclusion, we presented new radio images of the Galactic center and showed 44 compact radio sources 
within the E arm. 
Radio sources with near-IR counterparts are thought to be 
massive young stellar objects (Yusef-Zadeh et al. 2014) whereas radio sources without any infrared 
counterparts are likely  associated with proplyds around low mass stars. We also showed that 
star formation is taking place at the edge of an elongated neutral cloud, represented as a dark feature 
in a continuum image at 34 GHz as well as being traced by H$_2$ emission. 
The southern edge of the RDC traces the E arm of Sgr A West, 
photoionized by the strong radiation field of the Galactic center whereas the northern edge of the RDC 
signifies low mass star formation activity.  
Future mid-IR and millimeter observations should
conclusively test whether proplyds lie close to Sgr A* and
potentially measure  the disk mass of  Galactic center proplyd-like objects.

Acknowledgments: This work is partially supported by the grant AST-0807400 from the NSF,  DPO986386 from the 
Australian Research Council and 
the  European Research Council under the European Union's Seventh Framework Program (FP/2007-2013). 
This paper makes use of the following ALMA data: ADS/JAO.ALMA\#2011.0.00005.SV. ALMA is a 
partnership of ESO (representing its member states), NSF (USA) and NINS (Japan), together with NRC (Canada) and NSC and 
ASIAA (Taiwan), in cooperation with the Republic of Chile. The Joint ALMA Observatory is operated by ESO, AUI/NRAO and 
NAOJ. The research leading to these results has received funding from 
ERC Grant Agreement n. 614922.

\begin{figure}
\center
\includegraphics[scale=0.6,angle=0]{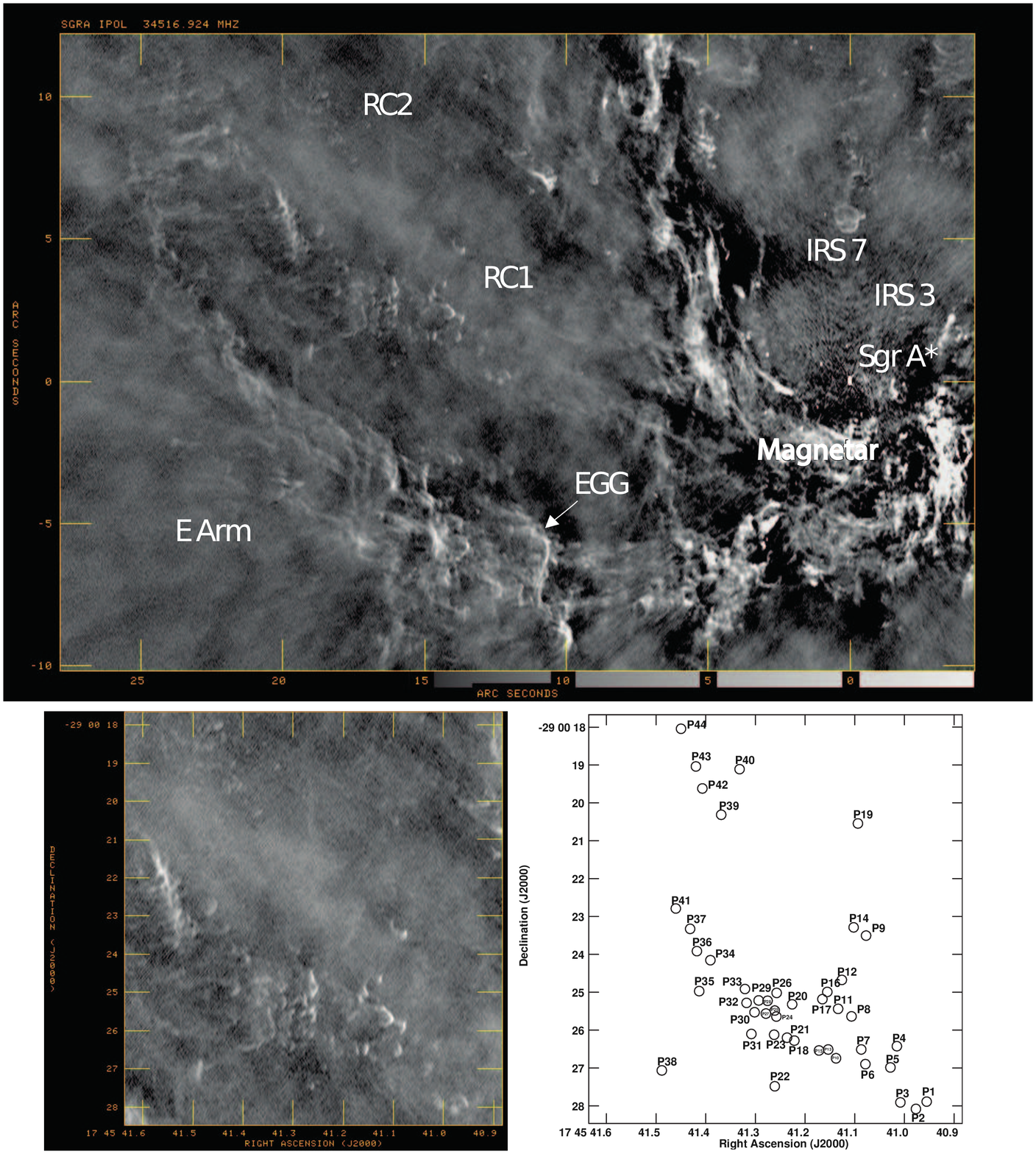}
\caption{
{\it (a)}
The prominent N and E arms of Sgr A West and the proplyd candidates (RC1 \&  RC2) and
an evaporating gaseous globule EGG candidate at 34 GHz. 
{\it (b)}
The same as (a) except a close up view of proplyd candidates.
 {\it (c)}
ID numbers of  of proplyd candidates (Table 1) are labeled over an identical  region 
shown in (b).  
}
\end{figure}

\begin{figure}
\center
\includegraphics[scale=0.6,angle=0]{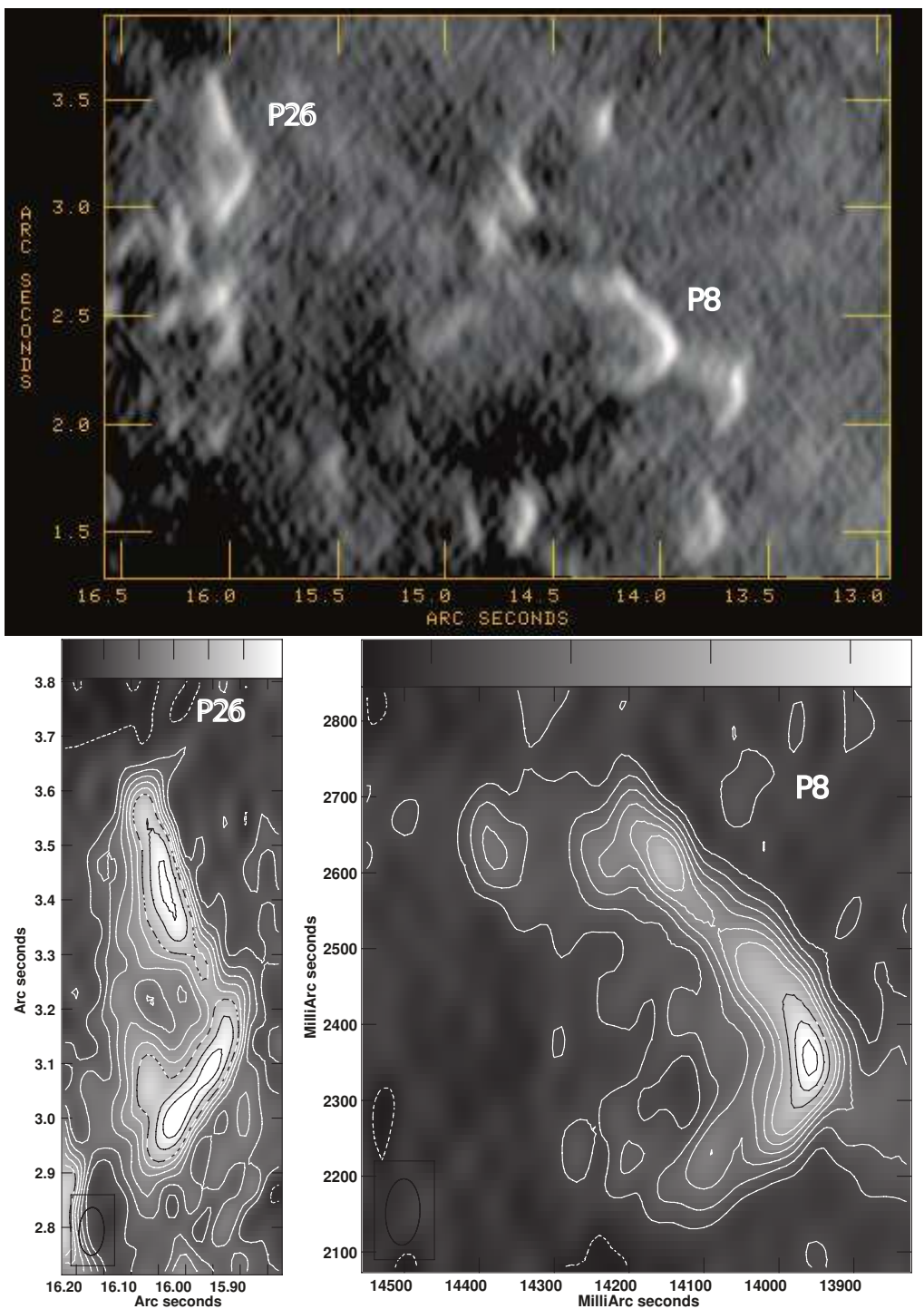}
\caption{
{\it (a)}
A 3.5$''\times2.5''$ of a 34 GHz image showing a concentration of proplyd candidates  offset by
15$''.2$ NE of  Sgr A*. 
{\it (b)}
Grayscale contours of P26 are set at -10, 10, 20,...60, 80,....120 $\mu$Jy beam$^{-1}$. 
{\it (c)}
Grayscale contours of P8 are set at -10, 10, 20,...60, 80 and  100 $\mu$Jy beam$^{-1}$. 
}
\end{figure}

\begin{figure}
\center
\includegraphics[scale=0.5,angle=0]{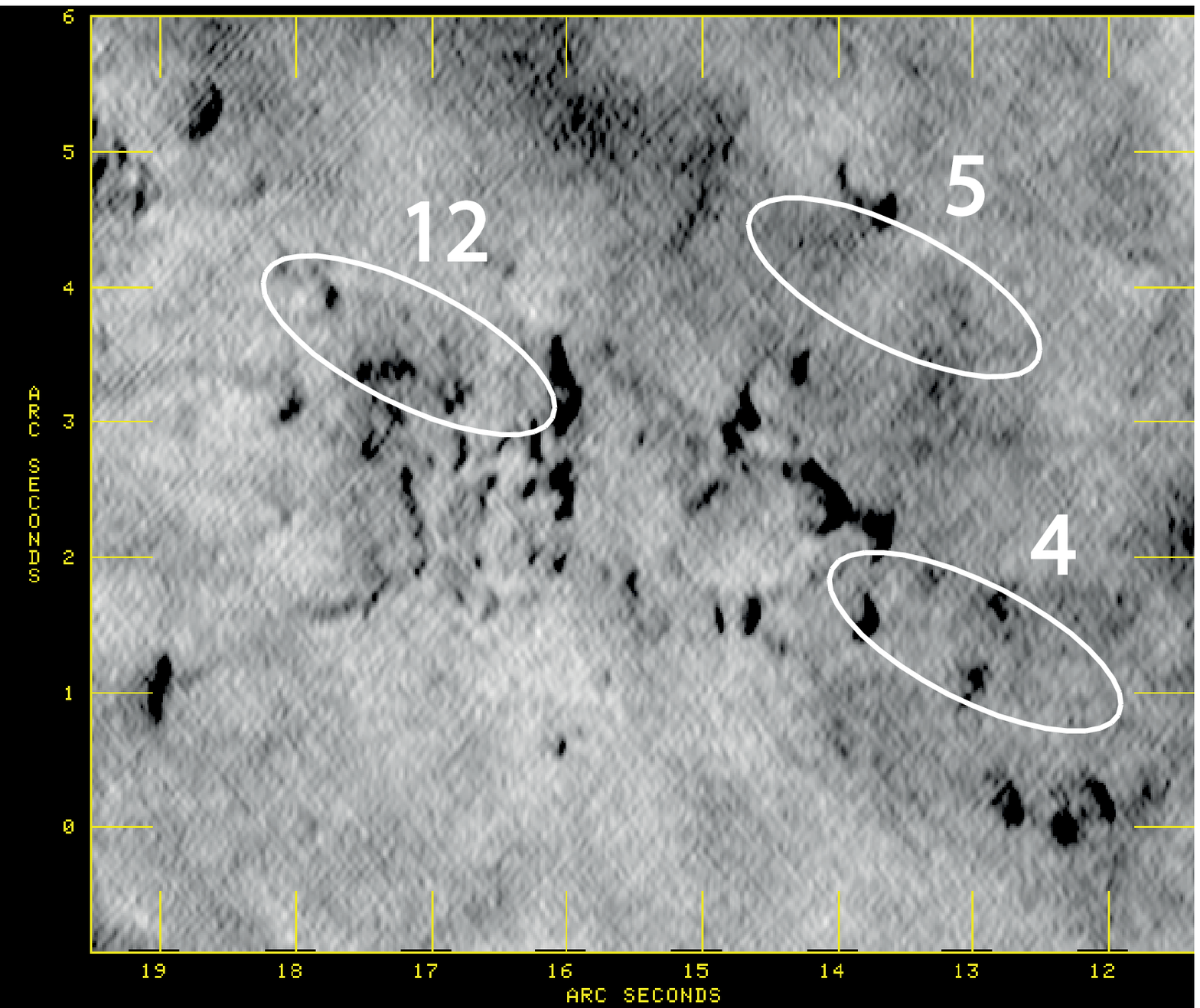}
\includegraphics[scale=0.5,angle=0]{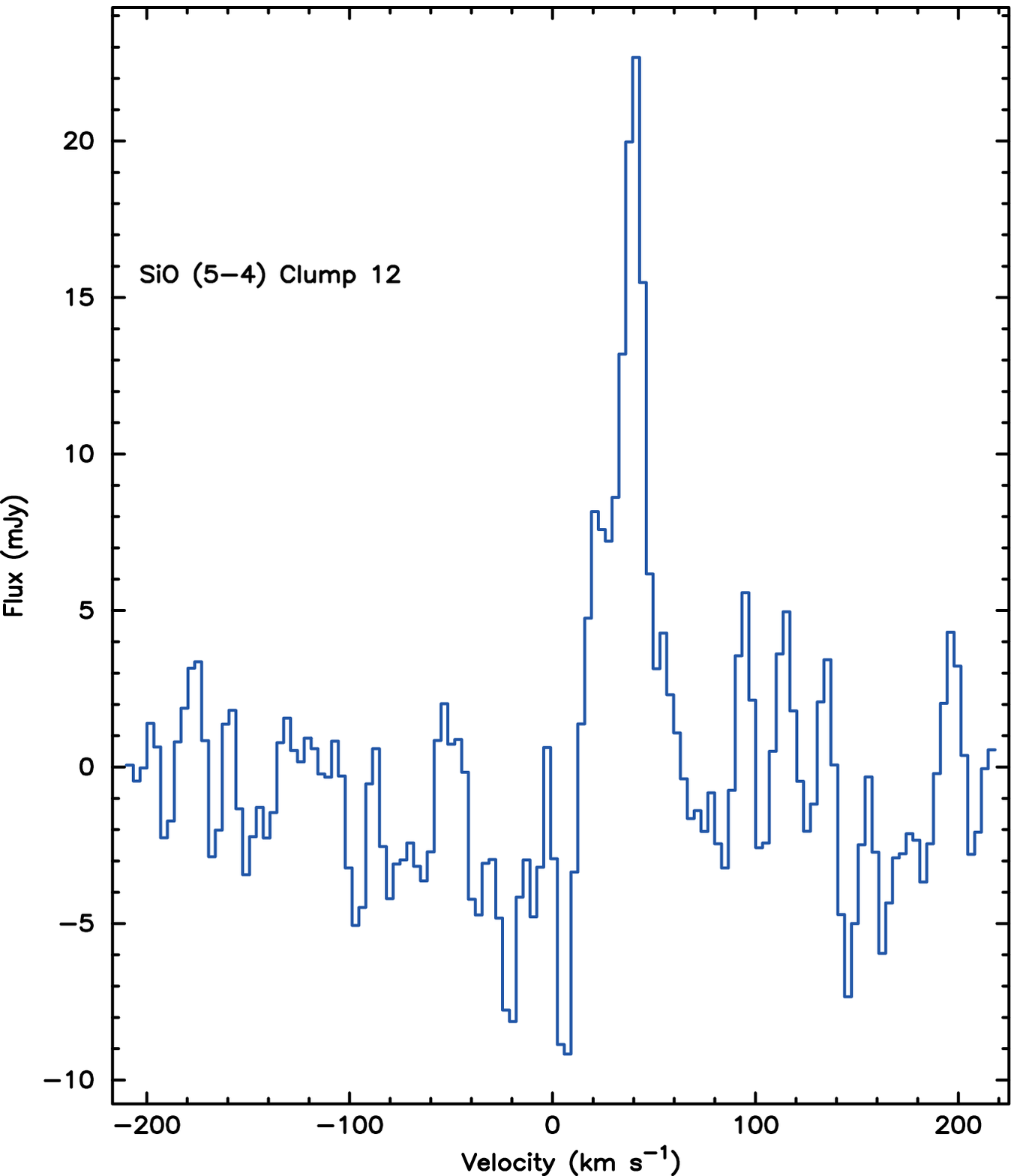}
\caption{
{\it (a)}
Three  ellipses corresponding to 
the location of three SiO (5-4) sources  4, 5 and 12
are drawn on a 34 GHz  grayscale image (reverse color) of proplyd-like objects. 
{\it (b)} 
The ALMA spectrum of source 12. 
 The size and position angle  of each ellipse corresponds 
 to  the spatial resolution of ALMA data (Yusef-Zadeh et al. 2013). 
}
\end{figure}

\begin{figure}
\center
\includegraphics[scale=0.6,angle=0]{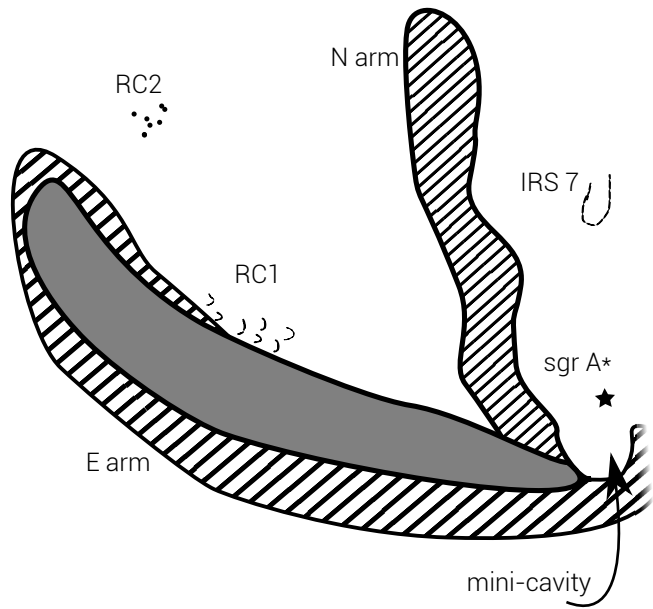}
\caption{
A schematic view of the region shown in Figure 1a. 
Shaded region coincides with radio dark clouds where hashed 
lines trace the distribution of ionized gas associated with the E and N arms 
of Sgr A West.
}
\end{figure}

\begin{deluxetable}{cccccccc}
\tablecaption{The  Parameters of Gaussian Fitted 34 GHz Sources}
\tabletypesize{\scriptsize}
\tablecolumns{10}
\tablewidth{0pt}
\setlength{\tabcolsep}{0.03in}
\tablehead{
\colhead{ID} & \colhead{RA (J2000)} & \colhead{Dec (J2000)} & \colhead{Dist. from Sgr A*} & \colhead{Pos. Accuracy} 
& \colhead{$ \theta_{a} \times \theta_{b} $ (PA)} & \colhead{Peak Intensity} & \colhead{Integrated Flux}\\
& ($17^{\rm h}45^{\rm m}$) & ($-29^{\circ}00^{\prime}$) & (\sl{arcsec}) & (\sl{mas}) 
& \sl{mas} $ \times $ \sl{mas} (\sl{deg}) & \sl{(mJy beam$^{-1}$)} & \sl{(mJy)}  
}
\startdata
  P1 & 40.9554 & 27.8883 & 12.03 & 17.15 & 242 $ \times $ 54 (27) & 0.077 $ \pm $ 0.013 & 0.376 $ \pm $ 0.077  \\
  P2 & 40.9770 & 28.0816 & 12.31 & 18.76 & 198 $ \times $ 155 (25) & 0.062 $ \pm $ 0.013 & 0.531 $ \pm $ 0.124  \\
  P3 & 41.0083 & 27.9076 & 12.73 & 27.53 & 360 $ \times $ 194 (19) & 0.053 $ \pm $ 0.010 & 0.961 $ \pm $ 0.184  \\
  P4 & 41.0153 & 26.4246 & 12.92 & 32.41 & 154 $ \times $ 86 (5) & 0.031 $ \pm $ 0.013 & 0.131 $ \pm $ 0.068  \\
  P5 & 41.0284 & 26.9837 & 13.03 & 30.20 & 339 $ \times $ 199 (159)  & 0.046 $ \pm $ 0.010 & 0.812 $ \pm $ 0.181  \\
  P6 & 41.0787 & 25.8970 & 13.82 & 8.61 & 169 $ \times $ 46 (172) & 0.126 $ \pm $ 0.013 & 0.383 $ \pm $ 0.053  \\
  P7 & 41.0871 & 26.5111 & 13.85 & 12.51 & 177 $ \times $ 50 (7) & 0.090 $ \pm $ 0.013 & 0.298 $ \pm $ 0.056  \\
  P8 & 41.1067 & 25.6379 & 14.22 & 17.45 & 477 $ \times $ 77 (38) & 0.112 $ \pm $ 0.012 & 1.340 $ \pm $ 0.153  \\
  P9 & 41.0773 & 23.5031 & 14.37 & 9.47 & 170 $ \times $ 88 (157) & 0.109 $ \pm $ 0.013 & 0.516 $ \pm $ 0.075  \\  
  P10 & 41.1378 & 26.7426 & 14.48 & 25.61 & 155 $ \times $ 109 (79) & 0.031 $ \pm $ 0.013 & 0.170 $ \pm $ 0.085  \\
  P11 & 41.1336 & 25.4366 & 14.61 & 15.48 & 103 $ \times $ 58 (40) & 0.046 $ \pm $ 0.014 & 0.123 $ \pm $ 0.048  \\
  P12 & 41.1260 & 24.6733 & 14.67 & 8.26 & 132 $ \times $ 42 (168) & 0.110 $ \pm $ 0.014 & 0.269 $ \pm $ 0.045  \\
  P13 & 41.1535 & 26.5116 & 14.71 & 10.70 & 212 $ \times $ 66 (168) & 0.119 $ \pm $ 0.013 & 0.539 $ \pm $ 0.072  \\
  P14 & 41.1024 & 23.2867 & 14.76 & 34.42 & 255 $ \times $ 57 (11) & 0.043 $ \pm $ 0.013 & 0.213 $ \pm $ 0.077  \\
  P15 & 41.1722 & 26.5353 & 14.95 & 16.37 & 186 $ \times $ 25 (11) & 0.071 $ \pm $ 0.014 & 0.195 $ \pm $ 0.049  \\
  P16 & 41.1555 & 24.9915 & 14.98 & 12.71 & 229 $ \times $ 58 (20) & 0.103 $ \pm $ 0.013 & 0.480 $ \pm $ 0.074  \\
  P17 & 41.1652 & 25.1813 & 15.06 & 13.94 & 189 $ \times $ 82 (9) & 0.083 $ \pm $ 0.013 & 0.400 $ \pm $ 0.075  \\
  P18 & 41.2218 & 26.2771 & 15.63 & 29.23 & 235 $ \times $ 91 (5) & 0.048 $ \pm $ 0.013 & 0.298 $ \pm $ 0.094  \\
  P19 & 41.0938 & 20.5442 & 15.76 & 37.64 & 208 $ \times $ 91 (179) & 0.033 $ \pm $ 0.013 & 0.187 $ \pm $ 0.086  \\
  P20 & 41.2261 & 25.3190 & 15.82 & 1.19 & 83 $ \times $ 38 (75) & 0.482 $ \pm $ 0.014 & 1.062 $ \pm $ 0.042  \\
  P21 & 41.2364 & 26.1998 & 15.83 & 27.52 & 105 $ \times $ 37 (12) & 0.029 $ \pm $ 0.014 & 0.059 $ \pm $ 0.039  \\
  P22 & 41.2612 & 27.4809 & 16.05 & 12.60 & 99 $ \times $ 34 (168) & 0.061 $ \pm $ 0.014 & 0.116 $ \pm $ 0.038  \\
  P23 & 41.2625 & 26.1203 & 16.08 & 14.97 & 113 $ \times $ 46 (140) & 0.049 $ \pm $ 0.014 & 0.126 $ \pm $ 0.046  \\
  P24 & 41.2580 & 25.6416 & 16.18 & 18.90 & 268 $ \times $ 66 (168) & 0.082 $ \pm $ 0.013 & 0.459 $ \pm $ 0.086  \\
  P25 & 41.2612 & 25.4816 & 16.25 & 6.93 & 64 $ \times $ 28 (139) & 0.088 $ \pm $ 0.014 & 0.140 $ \pm $ 0.034  \\
  P26 & 41.2573 & 25.0159 & 16.28 & 12.98 & 231 $ \times $ 27 (148) & 0.094 $ \pm $ 0.013 & 0.363 $ \pm $ 0.063  \\
  P27 & 41.2789 & 25.5619 & 16.47 & 11.17 & 154 $ \times $ 36 (135) & 0.073 $ \pm $ 0.013 & 0.230 $ \pm $ 0.054  \\
  P28 & 41.2756 & 25.2243 & 16.48 & 20.27 & 259 $ \times $ 37 (4) & 0.076 $ \pm $ 0.013 & 0.302 $ \pm $ 0.065  \\
  P29 & 41.2935 & 25.2169 & 16.71 & 23.66 & 265 $ \times $ 51 (136) & 0.050 $ \pm $ 0.013 & 0.285 $ \pm $ 0.086  \\
  P30 & 41.3018 & 25.5289 & 16.77 & 13.65 & 144 $ \times $ 63 (143) & 0.062 $ \pm $ 0.013 & 0.217 $ \pm $ 0.058  \\
  P31 & 41.3083 & 26.1015 & 16.78 & 13.89 & 80 $ \times $ 23 (28) & 0.049 $ \pm $ 0.014 & 0.081 $ \pm $ 0.034  \\  
  P32 & 41.3180 & 25.2808 & 17.02 & 28.45 & 254 $ \times $ 78 (177) & 0.053 $ \pm $ 0.013 & 0.313 $ \pm $ 0.090  \\
  P33 & 41.3212 & 24.9135 & 17.12 & 26.66 & 229 $ \times $ 57 (134) & 0.039 $ \pm $ 0.013 & 0.205 $ \pm $ 0.081  \\
  P34 & 41.3907 & 24.1522 & 18.17 & 18.15 & 174 $ \times $ 83 (11) & 0.060 $ \pm $ 0.013 & 0.272 $ \pm $ 0.072  \\
  P35 & 41.4134 & 24.9705 & 18.30 & 23.68 & 232 $ \times $ 107 (150) & 0.053 $ \pm $ 0.013 & 0.384 $ \pm $ 0.107  \\
  P36 & 41.4180 & 23.9140 & 18.57 & 32.84 & 188 $ \times $ 87 (175) & 0.035 $ \pm $ 0.013 & 0.177 $ \pm $ 0.078  \\
  P37 & 41.4316 & 23.3262 & 18.88 & 42.84 & 218 $ \times $ 125 (162) & 0.030 $ \pm $ 0.013 & 0.226 $ \pm $ 0.112  \\
  P38 & 41.4885 & 27.0644 & 19.05 & 28.39 & 441 $ \times $ 145 (168) & 0.066 $ \pm $ 0.010 & 1.105 $ \pm $ 0.177  \\
  P39 & 41.3689 & 20.3132 & 19.10 & 21.94 & 179 $ \times $ 52 (119) & 0.034 $ \pm $ 0.013 & 0.146 $ \pm $ 0.069  \\
  P40\tablenotemark{a} & 41.3320 & 19.1091 & 19.19 & 17.19 & 121 $ \times $ 1 (138) & 0.043 $ \pm $ 0.014 & 0.095 $ \pm $ 0.042  \\
  P41 & 41.4604 & 22.7882 & 19.39 & 36.96 & 385 $ \times $ 159 (156) & 0.043 $ \pm $ 0.010 & 0.695 $ \pm $ 0.174 \\
  P42\tablenotemark{a} & 41.4068 & 19.6192 & 19.84 & 20.02 & 144 $ \times $ 0 (142) & 0.042 $ \pm $ 0.014 & 0.090 $ \pm $ 0.041  \\
  P43 & 41.4199 & 19.0363 & 20.25 & 22.69 & 140 $ \times $ 17 (143) & 0.036 $ \pm $ 0.014 & 0.090 $ \pm $ 0.045  \\
  P44 & 41.4499 & 18.0418 & 21.06 & 18.59 & 227 $ \times $ 122 (119) & 0.053 $ \pm $ 0.013 & 0.439 $ \pm $ 0.120  \\
\enddata
\tablenotetext{a}{spatially unresolved in one direction}
\end{deluxetable}

\end{document}